\documentclass[11pt]{article}
\usepackage{amsfonts,amsmath,amssymb,amsfonts,graphicx,hyperref}
\begin{document}

\title{New Financial Research Program:\\ General Option-Price Wave Modeling}
\author{Vladimir G. Ivancevic \\
{\small Defence Science \& Technology Organisation, Australia}}
\date{}
\maketitle

\begin{abstract}
Recently, a novel adaptive wave model for financial option pricing has been proposed in the form of adaptive nonlinear Schr\"{o}dinger (NLS) equation \cite{IvCogComp}, as a high-complexity alternative to the linear Black--Scholes--Merton model \cite{BS,Merton}. Its quantum-mechanical basis has been elaborated in \cite{IvCejp}. Both the solitary and shock-wave solutions of the nonlinear model, as well as its linear (periodic) quantum simplification are shown to successfully fit the Black--Scholes data, and define the financial Greeks. This initial wave model (called the Ivancevic option pricing model) has been further extended in \cite{Yan}, by providing the new NLS solutions in the form of rogue waves (one-rogon and two-rogon solutions).
In this letter, I propose a new financial research program, with a goal to develop a general wave-type model for realistic option-pricing prediction and control. \\

\end{abstract}

\section{Introduction}

Recently, a novel adaptive wave model for financial option pricing has been proposed in the form of adaptive nonlinear Schr\"{o}dinger (NLS) equation \cite{IvCogComp}, as a high-complexity alternative to the linear Black--Scholes-Merton model \cite{BS,Merton}. The new option-pricing model, representing a \emph{controlled Brownian motion}, comes in two flavors: (i) for the case of constant volatility, it is defined by a single adaptive NLS equation, while for the case of stochastic volatility, it is defined by an adaptive Manakov system of two coupled NLS equations.

The adaptive wave model with stock price $s,$ volatility $\sigma $ and interest rate $r$ has been formally defined in \cite{IvCogComp,IvCejp} as a complex-valued, focusing NLS equation, defining the
time-dependent \emph{option--price wave function} $\psi =\psi (s,t)$, whose
absolute square\ $|\psi (s,t)|^{2}$ represents the probability density
function (PDF) for the option price in terms of the stock price and time. In
natural quantum units, this adaptive (1+1)--NLS equation (called the \emph{Ivancevic option pricing model}):\footnote{Physically, the NLS equation (\ref{nlsGen}) describes a
nonlinear wave (e.g., in optical fibers or Bose-Einstein condensates) defined by the complex-valued wave function $\psi
(s,t)$ of real space and time parameters. In the present context, the
space-like variable $s$ denotes the stock (asset) price.}
\begin{equation}
\mathrm{i}\partial _{t}\psi =-\frac{1}{2}\sigma \partial _{ss}\psi -\beta
|\psi |^{2}\psi ,\qquad (\mathrm{i}=\sqrt{-1}),   \label{nlsGen}
\end{equation}
where $\beta=\beta (r,w) $ denotes the adaptive market-heat potential, so the term $V(\psi )=-\beta |\psi |^{2}$ represents the $%
\psi -$dependent potential field. In the simplest
nonadaptive scenario $\beta $ is equal to the interest rate $r$, while in
the adaptive case it depends on the set of adjustable synaptic weights $\{w^i_j\}$ as:~
$
\beta (r,w)=r\sum_{i=1}^{n}w_{1}^{i}\,\text{erf}\left( \frac{w_{2}^{i}s}{%
w_{3}^{i}}\right).
$

\section{Building the Wave Model for Option Pricing Evolution}

The development of the general wave option-pricing model can be summarized as follows.

Firstly, in the case of low interest-rate $r\ll 1$, we have $\beta (r)\ll 1$, so $%
V(\psi )\rightarrow 0,$ and therefore equation (\ref{nlsGen}) can be
approximated by a quantum-like \emph{option wave packet,} `physically'
associated with a free quantum particle of unit mass. This linear wave
packet is a solution of the \emph{linear Schr\"{o}dinger equation} (with
zero potential energy, Hamiltonian operator $\hat{H}$ and volatility $\sigma
$ playing the role similar to the Planck constant) \cite{IvCejp}:
\begin{equation}
\mathrm{i}\sigma \partial _{t}\psi =\hat{H}\psi ,\qquad \text{where}\qquad
\hat{H}=-\frac{\sigma ^{2}}{2}\partial _{ss}.  \label{sch1}
\end{equation}

The general solution to (\ref{sch1}) is usually formulated as a linear combination
of de Broglie-type \emph{option plane-waves} with the wave number $k$, linear momentum $p=\sigma k,$
wavelength $\lambda _{k}=2\pi /k,$\ angular frequency $\omega _{k}=\sigma
k^{2}/2,$ and oscillation period $T_{k}=2\pi /\omega _{k}=4\pi /\sigma k^{2}$ -- all comprising the \emph{option wave-packet}:
\begin{equation}
\psi_{\rm packet}(s,t)=\sum_{i=0}^{n}c_{i}A\mathrm{e}^{\mathrm{i}(ks-\omega _{k}t)},\qquad (\text{with}\ n\in
\Bbb{N}),  \label{w-pack}
\end{equation}
where $A$ is the amplitude of the option wave, the angle $(ks-\omega _{k}t)=(ks-{%
\frac{\sigma k^{2}}{2}}t)$\ represents the phase of the wave $\psi _{k}$
with the \emph{phase velocity:} $v_{k}=\omega _{k}/k=\sigma k/2.$ In addition, the \emph{group velocity} of an option wave-packet is given by: $\ v_{g}=d\omega
_{k}/dk.$ It is related to the phase velocity $v_{k}$ of a plane wave as: $%
v_{g}=v_{k}-\lambda _{k}dv_{k}/d\lambda _{k}.$ Closely related is the \emph{center} of the option wave-packet (the point of maximum amplitude), given by: $%
s=td\omega _{k}/dk.$ The option wave-packet has been used used in \cite{IvCejp} to successfully fit the  Black--Scholes call and put options data.

Secondly, the NLS equation (\ref{nlsGen}) has been exactly solved in \cite{IvCogComp} using the power series expansion method of Jacobi elliptic functions. Out of a series of solutions, the two most important ones (from dynamical perspective) are:
\begin{enumerate}
  \item The envelope \emph{shock-wave} solution (or, `dark soliton'):\footnote{%
A shock wave is a type of fast-propagating nonlinear disturbance that
carries energy and can propagate through a medium (or, field). It is
characterized by an abrupt, nearly discontinuous change in the
characteristics of the medium. The energy of a shock wave dissipates
relatively quickly with distance and its entropy increases. On the other
hand, a soliton is a self-reinforcing nonlinear solitary wave packet that
maintains its shape while it travels at constant speed. It is caused by a
cancelation of nonlinear and dispersive effects in the medium (or, field).}
\begin{equation}
\psi_{\textrm{shock}}(s,t) =\pm \sqrt{\frac{-\sigma }{\beta}}\,\mathrm{\tanh }%
(s-\sigma kt)\,\mathrm{e}^{\mathrm{i}[ks-\frac{1}{2}\sigma
t(2+k^{2})]},\qquad\text{and}  \label{tanh1}
\end{equation}
  \item The envelope \emph{solitary-wave} solution (or, `bright soliton'):
\begin{equation}
\psi_{\textrm{soliton}}(s,t) =\pm \sqrt{\frac{\sigma }{\beta}}\,\mathrm{sech}%
(s-\sigma kt)\,\mathrm{e}^{\mathrm{i}[ks-\frac{1}{2}\sigma
t(k^{2}-1)]}.  \label{sech1}
\end{equation}
\end{enumerate}

The adaptive NLS--PDFs of the combined shock-wave type (\ref{tanh1}) and soliton type (\ref{sech1}) have been used in \cite{IvCogComp} to successfully fit the Black--Scholes call and put options data. Besides, the
adaptive NLS--based Greeks (Delta, Rho, Vega, Theta and Gamma) have been
defined, as partial derivatives of the shock-wave solution (\ref{tanh1}).

Thirdly, two new wave-solutions of the NLS equation (\ref{nlsGen}) have been provided in \cite{Yan}, in the form of rogue waves,\footnote{Rogue waves are also known as
{\it freak waves}, {\it monster waves}, {\it killer waves}, {\it
giant waves} and {\it extreme waves}. They are found in various media, including optical fibers \cite{Solli}. The basic rogue wave solution was first presented by Peregrine~\cite{Peregrine} to describe the phenomenon known as \emph{Peregrine soliton} (or Peregrine breather).} using the deformed Darboux transformation method developed in \cite{Akhmediev}.
\begin{enumerate}
  \item The \emph{one-rogon} solution:
\begin{equation}
\psi_{1\rm rogon}(s,t)=\alpha\sqrt{\frac{\sigma}{2\beta}}\left[1-\frac{4(1+\sigma\alpha^2t)}
  {1+2\alpha^2(s-\sigma kt)^2+\sigma^2\alpha^4t^2}\right]\,{\rm e}^{{\rm i}[ks+\sigma/2(\alpha^2-k^2)t]},
  \quad \sigma\beta>0, \label{rog1}
\end{equation}
where $\alpha$ and $k$ denote the scaling and gauge.
  \item The \emph{two-rogon} solution:
\begin{equation}
\psi_{2\rm rogon}(s,t)=\alpha\sqrt{\frac{\sigma}{2\beta}}
 \left[1+\frac{P_2(s,t)+{\rm i}Q_2(s,t)}{R_2(s,t)}\right]\,{\rm e}^{{\rm i}[ks+\sigma/2(\alpha^2-k^2)t]}, \quad \sigma\beta>0,  \label{rog2}
\end{equation}
where $P_2,Q_2,R_2$ are certain polynomial functions of $s$ and $t$.
\end{enumerate}

\section{New Financial Research Program}

I propose a new financial research program as follows.

Firstly, define the \emph{general adaptive wave model for option pricing evolution} as a (linear) combination of the previously defined particular solutions to the adaptive NLS-equation (\ref{nlsGen}). The five wave-components of this general model are:
\begin{enumerate}
  \item the linear wave packet $\psi_{\rm packet}(s,t)$, given by (\ref{w-pack});
  \item the shock-wave $\psi_{\textrm{shock}}(s,t)$, given by (\ref{tanh1});
  \item the soliton $\psi_{\textrm{soliton}}(s,t)$, given by (\ref{sech1});
  \item the one-rogon $\psi_{1\rm rogon}(s,t)$, given by (\ref{rog1}); ~and
  \item the two rogon $\psi_{2\rm rogon}(s,t)$ (\ref{rog2}).
\end{enumerate}
Formally, the general adaptive wave model is defined by:
\begin{eqnarray}
\psi_{\rm general}(s,t) &=& A_1\sum_{i=0}^{n}c_{i}\mathrm{e}^{\mathrm{i}(ks-\omega _{k}t)} \label{genWave} \\
&\pm&  A_2\sqrt{\frac{-\sigma }{\beta}}\,\mathrm{\tanh }(s-\sigma kt)\,\mathrm{e}^{\mathrm{i}[ks-\frac{1}{2}\sigma t(2+k^{2})]} \notag\\
&\pm&  A_3\sqrt{\frac{\sigma }{\beta}}\,\mathrm{sech}(s-\sigma kt)\,\mathrm{e}^{\mathrm{i}[ks-\frac{1}{2}\sigma
t(k^{2}-1)]} \notag\\
&+& A_4\alpha\sqrt{\frac{\sigma}{2\beta}}\left[1-\frac{4(1+\sigma\alpha^2t)}
  {1+2\alpha^2(s-\sigma kt)^2+\sigma^2\alpha^4t^2}\right]\,{\rm e}^{{\rm i}[ks+\sigma/2(\alpha^2-k^2)t]} \notag\\
&+& A_5\alpha\sqrt{\frac{\sigma}{2\beta}}
 \left[1+\frac{P_2(s,t)+{\rm i}Q_2(s,t)}{R_2(s,t)}\right]\,{\rm e}^{{\rm i}[ks+\sigma/2(\alpha^2-k^2)t]}, \notag
\end{eqnarray}
where $A_i,~(i=1,...,5)$ denote adaptive amplitudes of the five waves, while the other parameters are defined in the previous section.

Secondly, we need to find the most representative financial index or contemporary markets data that clearly show in their evolution both the \emph{efficient markets hypothesis} \cite{Lo1} and \emph{adaptive markets hypothesis} \cite{Lo2}. Once we find such a representative data, we need to fit it using our general wave model (\ref{genWave}) and the powerful Levenberg-Marquardt fitting algorithm.
I remark here that, based on my empirical experience, the general wave model (\ref{genWave}) is capable of fitting any financial data, provided we use appropriate number of fitting coefficients (see \cite{IvCogComp,IvCejp} for technical details).

Once we have successfully fitted the most representative market data we will have a model that can be used for \emph{prediction} of many possible outcomes of the current \emph{global financial storm}.

\end{document}